\documentclass[12pt]{iopart}
\usepackage[dvips]{graphicx, color} 
\begin{document}
\title[Electrical transport of  Heusler compounds CoFeTiSn and CoFeVGa] {Unconventional transport behaviour in the quaternary Heusler compounds CoFeTiSn and CoFeVGa}
\author{Snehashish Chatterjee$^{a}$, Subarna Das$^{c}$, S. Pramanick$^{b}$, S. Chatterjee$^{b}$, S. Giri$^{a}$, Aritra Banerjee$^{c}$, S. Majumdar$^{a*}$}
\address{Department of Solid State Physics, Indian Association for the Cultivation of Science, 2A \& 2B Raja S. C. Mullick Road, Jadavpur, Kolkata 700 032, India}
\address{$^{b}$UGC-DAE Consortium for Scientific Research, Kolkata Centre, Sector III, LB-8, Salt Lake, Kolkata 700 098, INDIA} 
\address{$^{c}$Department of Physics, University of Calcutta, Kolkata 700 009, INDIA} 

\ead{$^{*}$sspsm2@iacs.res.in}

\begin{abstract}
We report here the electrical transport and magnetic properties of the newly synthesized quaternary Heusler compound  CoFeTiSn and CoFeVGa. We observe a striking change in the electronic transport properties of CoFeTiSn as the system undergoes the paramagnetic to ferromagnetic transition. While the sample shows an activated semiconducting behaviour in the paramagnetic phase, it turns abruptly to a metallic phase with the onset of ferromagnetic transition. We have compared the system with other Hesuler compounds showing similar anomaly in transport, and it appears that CoFeTiSn has much similarities with the Fe$_2$VAl compound having pseudogap in the paramagnetic phase. In sharp contrast, CoFeVGa shows a predominantly semiconducting behaviour down to 90 K, below which it shows a window of metallic region. Both the compositions show negative Seebeck coefficient varying linearly with temperature. The value of the Seebeck coefficient of CoFeTiSn is comparable to  that of many Heusler alloys identified as potential thermoelectric materials.        
\end{abstract}

\maketitle

\section{Introduction}
Recently, transition metal based Heusler alloys~\cite{heusler} have attracted considerable attention due to their diverse and intriguing electronic properties, which include  half-metallic ferromagnetism, magnetic shape memory effect,  Kondo behaviour, large thermo-electric effect, unconventional superconductivity and so on~\cite{parkin,krenke,kainuma,kondo,thermo,mpd2pb}. They represent a class of ternary intermetallic compounds having general formula X$_2$YZ where X and Y are transition metals whereas Z is a nonmagnetic {\it sp} element. These compositions crystallize in a cubic L2$_1$ structure  with space group $Fm\bar{3}m$. However, a different structural symmetry is obtained when one of the X atoms is replaced by a different transition metal atom. This results in equiatomic quaternary Heusler compounds, XX$^{\prime}$YZ, which retains its cubic symmetry, but have a different space group ($F\bar{4}3m$). 
\par
The study of these quaternary Heusler compounds is an active area of research because of their possible application as a spin-polarized material in the  field of  spintronics~\cite{suresh}. Apart from their half metallic character, some of the quaternary compounds are identified as spin-gapless-semiconductors (SGSs)~\cite{cofemnsi}. SGS is a subgroup of half metallic ferromagnet (HMF), where the minority spin-down electrons has a gap at the Fermi level, while the majority spin-up electrons are semiconducting with a zero gap. The use of SGS in the field of spintronics can be quite rewarding considering full spin-polarization of both electrons and holes~\cite{ozdogan}. In last one decade several HMF and SGS have been identified among quaternary Heusler compounds, based on theoretical calculations as well as on experimental studies.

\par
XX$^{\prime}$YZ compounds can be synthesized with variety of atoms at the X, X$^{\prime}$, Y and Z sites, which provides us a flexibility to tune the electronic structure over a wide range starting from a band insulator to a metal. For full Heusler compounds crystallizing in L2$_1$ structure, it is found that half-metallicity prevails if the compound obeys a case of Slater-Pauling rule, $M_t = Z_t -$ 24 (`rule of 24'), where $Z_t$ is the total number of valence electrons in the compound, and $M_t$ is the total magnetic moment (in Bohr magneton, $\mu_B$) per formula unit (f.u.)~\cite{sp}. For C1$_b$ type half Heusler alloys, a similar `rule of 18' ($M_t = Z_t -$ 18) is observed for half metallicity. For the presently studied quaternary compounds, the `rule of 24' is often found to be valid~\cite{hm}. It has been argued that quaternary compounds with $Z_t$ = 21 and 26 are good candidates to be SGS~\cite{ozdogan}. 

\par
In the present work, we focus on the experimental investigation on two such compounds, namely CoFeTiSn and CoFeVGa. Many of the CoFeYZ alloys are found to be possible candidates for HMF or SGS, which include CoFeTiZ (Z = Si, Ge, Sn)~\cite{cofetisn}, CoFeCrGe~\cite{cofecrge}, CoFeCrSi and CoFeCrGe~\cite{jin}, CoFeCrZ (Z = P, As and Sb)~\cite{cofecrz}, CoFeMnSi~\cite{cofemnsi} and so on. Both the compounds have valence electron concentration, $Z_t$ = 25. First principle electronic structure calculations indicate that the majority electrons are metallic in CoFeTiSn, while the Fermi level touches the top of the minority valence band, and the material is described as a quasi-half metallic ferrimagnet with high degree of spin polarization~\cite{cofetisn}.  CoFeVGa is also found to be a nearly half metal on the basis of first principle calculations~\cite{lun}. In the present work, we find that CoFeTiSn is a ferromagnet with Curie temperature ($T_C$) close to room temperature, along with the observation of an unusual semiconductor to metal transition on cooling at $T_C$. CoFeVGa, on the other hand, shows a predominantly semiconducting behaviour with an magnetic transition at $T_N$ = 20 K.
                     
\section{Experimental Details}
Polycrystalline samples of CoFeTiSn and CoFeVGa were prepared by standard arc melting technique. Next, the arc molten ingots were sealed in vacuum in a quartz capsule and annealed at 800$^{\circ}$ C for 3 days followed by normal furnace cooling down to room temperature. The crystallographic structure was investigated by X-ray powder diffraction (XRD) and was analyzed using the Rietveld refinement technique with MAUD program package~\cite{maud}. Magnetic measurements were carried out by using a commercial Quantum Design SQUID magnetometer (MPMS XL Ever Cool model). The resistivity ($\rho$) was measured by four probe method on a homemade setup fitted in a nitrogen cryostat (between 77 and 400 K) as well as on cryogen-free high magnetic field system (Cryogenic Ltd., U.K.) between 5-300 K.  Thermoelectric measurements were performed using standard  differential  technique  in  the  range of 20 K-300 K.  

\begin{figure}[t]
\centering
\includegraphics[width = 15 cm]{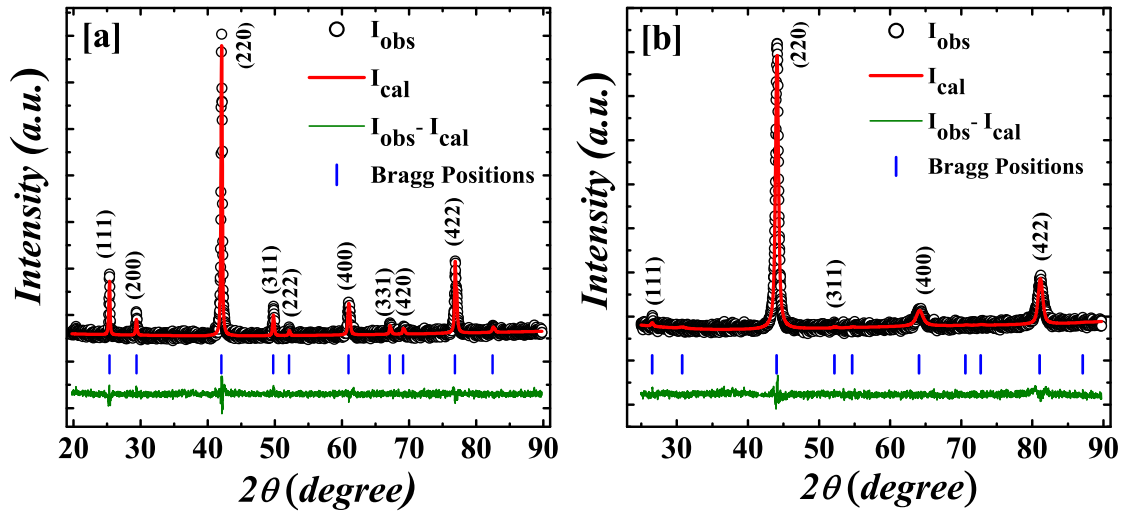}
\caption {Room temperature XRD pattern of (a) CoFeTiSn and (b) CoFeVGa (data points) recorded on a Bruker AXS diffractometer (D8 advance) using Cu K$_{\alpha}$ radiation along with the Rietveld refinement curves (solid lines).}
\label{xrd}
\end{figure}

\section{Results}
\subsection{Powder X-ray Diffraction}
Fig.~\ref{xrd} depicts powder XRD pattern of the studied samples measured at room temperature. Observation of superlattice reflections in the XRD pattern  indicates a well-ordered cubic Heusler  structure for CoFeTiSn. Rietveld refinement of the pattern confirms the formation of LiMgPdSn-type Y structure where the atoms occupy Wyckoff positions 4a, 4b, 4c and   4d. The atomic positions are as follows: Co(3/4, 3/4,3/4), Fe(1/4,1/4,1/4), Ti(1/2,1/2,1/2) and Sn(0,0,0). In case of CoFeVGa, the superlattice reflection (111) is  less intense than that observed in CoFeTiSn.  The weaker  (111) reflections might be due to similar scattering amplitudes of Co, Fe,V and Ga atoms. As pointed out by  Bainsla {\it et al.}~\cite{bainsla}, when the constituent elements of the sample are from the same period of the periodic table, it is difficult to identify the exact crystal structure. Nevertheless, good refinements of the data indicates that both CoFeTiSn and CoFeVGa form an ordered Y-type structure. The refined cubic lattice parameters are found to be 6.069 \AA~ for CoFeTiSn and 5.806 \AA~ for CoFeVGa which are quite close to the theoretical predictions  (6.09 \AA~ and 5.73 \AA ~respectively~\cite{cofetisn,lun}).

\begin{figure}[t]
\centering
\includegraphics[width = 7.8 cm]{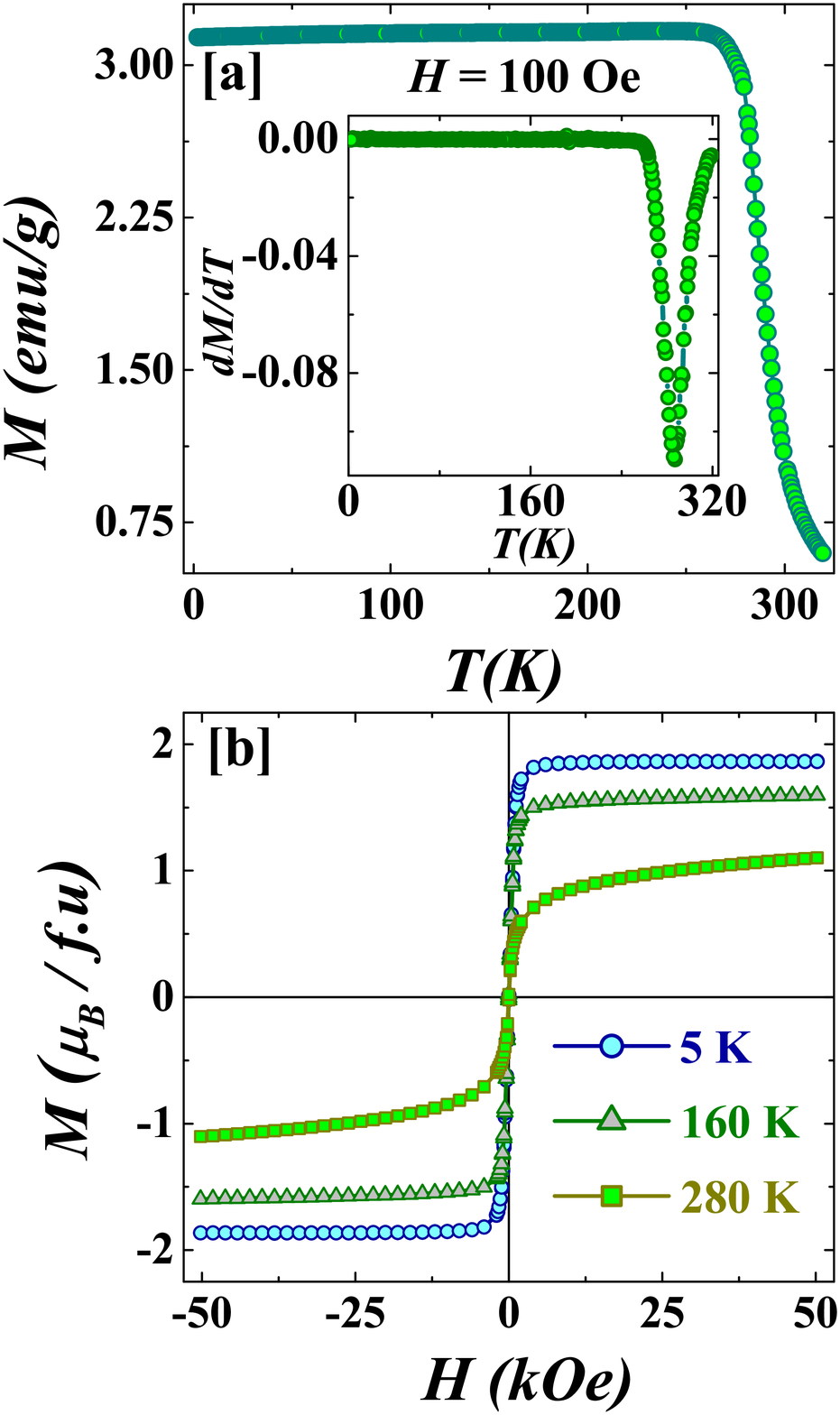}
\caption {(a) shows magnetization versus temperature  curve for $H$ = 100 Oe measured in the field-cooled condition for CoFeTiSn. The inset of (a) shows $dM/dT$ as a function of temperature. (b) shows isothermal $M$-$H$ curves measured at different constant temperatures.}
\label{mag}
\end{figure}

\subsection{Magnetization}
We  presented the  temperature ($T$) variation of magnetization ($M$) in the zero-field-cooled condition  for $H$ = 100 Oe of magnetic field  between 320 to 2 K for CoFeTiSn in the main panel of fig.~\ref{mag}(a). $M(T)$ data are  characterized by  a sharp rise just below room temperature, indicating a paramagnetic (PM) to ferromagnetic (FM) transition on cooling. Below the transition, $M$ shows an almost constant behaviour down to the lowest temperature of measurement. $T$-dependence of $dM/dT$ is shown in the inset of fig.~\ref{mag}(a), and the estimated $T_C$ from the minimum of the $dM/dT$ plot is found to be 287 K. The isothermal $M$ vs. $H$ curves recorded at different temperatures  are plotted in fig.~\ref{mag}(b), which reaffirms the FM nature of the sample. The isotherm recorded at 5 K attains its technical saturation at an applied field of 2.5 kOe, and the saturation moment is close to 2 $\mu_B$/f.u. This clearly indicates that although it shows an almost integer value of saturation moment, the `$Z_t -$ 24' Slater-Pauling rule is not obeyed here. The sample shows vanishingly small coercive field, indicating very soft FM character.              

\par
\begin{figure}[t]
\centering
\includegraphics[width = 7.8 cm]{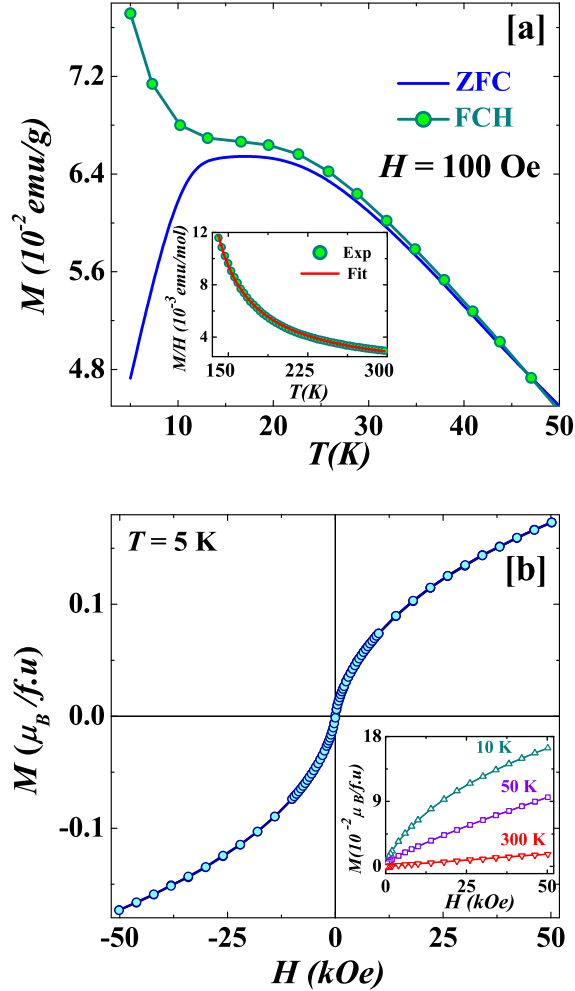}
\caption {(a) shows magnetization as a function of temperature for $H$ = 100 Oe recorded both in the field-cooled heating (FCH) and zero-field-cooled (ZFC) conditions for CoFeVGa. Inset of (a) shows the $M/H$ versus $T$ data  plotted between 140 to 300 K. The solid line represents Curie-Weiss fit to the data. (b) shows isothermal $M$-$H$ curves measured at different constant temperatures.}
\label{mag1}
\end{figure}

 
The $M$ vs. $T$ curve [main panel of fig.~\ref{mag1}(a))] for CoFeVGa sample is recorded in both field-cooled heating (FCH) and zero-field-cooling (ZFC) conditions under $H$ = 100 Oe. The ZFC data show a clear peak at $T_N$ = 20 K, mimicking the behaviour of an antiferromagnetic (AFM)-like transition. The ZFC and FC curves also bifurcate from  above $T_N$. The high temperature data (140-300 K) can be well fitted [see the inset of fig.~\ref{mag1} (a)] with a modified Curie-Weiss law, $M/H = C/(T-\theta) + \chi_0 $, where $C$ is the Curie constant, $\theta$ is the paramagnetic Curie temperature and $\chi_0$ is the $T$-independent susceptibility. The good quality of fitting indicates that the system is in a paramagnetic phase above $T_N$ with localized moment. From the fitting, we obtain the paramagnetic moment, $\mu_{eff}$ = 1.75 $\mu_B$, and  $\theta$ = 104 K. The value of $\chi_0$ ($\sim$ 10$^{-5}$ emu/mol) is found to be small but positive, indicating that it originates from Pauli paramagnetism. The positive $\theta$ value indicates predominant FM correlation in the system.            
\par
The isothermal magnetization measurement at 5 K [main panel of fig.~\ref{mag1}(b)] reveals a non-linear behaviour with curvature at higher fields. This is not expected for a purely AFM system. The coercive field for CoFeVGa  is also vanishingly small indicating the absence of any loop.
The inset of  fig.~\ref{mag1} (b)  represents $M(H)$ curves measured at $T$ = 10, 50 and 300 K for 0 to 50 kOe field cycling. The 10 K isotherm shows clear signature of curvature, while the other isotherms recorded at higher temperatures  are found to be linear in nature. Evidently, the observed value of $M$ at 50 kOe in CoFeVGa is about two order of magnitude less than that of  CoFeTiSn. 

\begin{figure}[b]
\centering
\includegraphics[width = 7.8 cm]{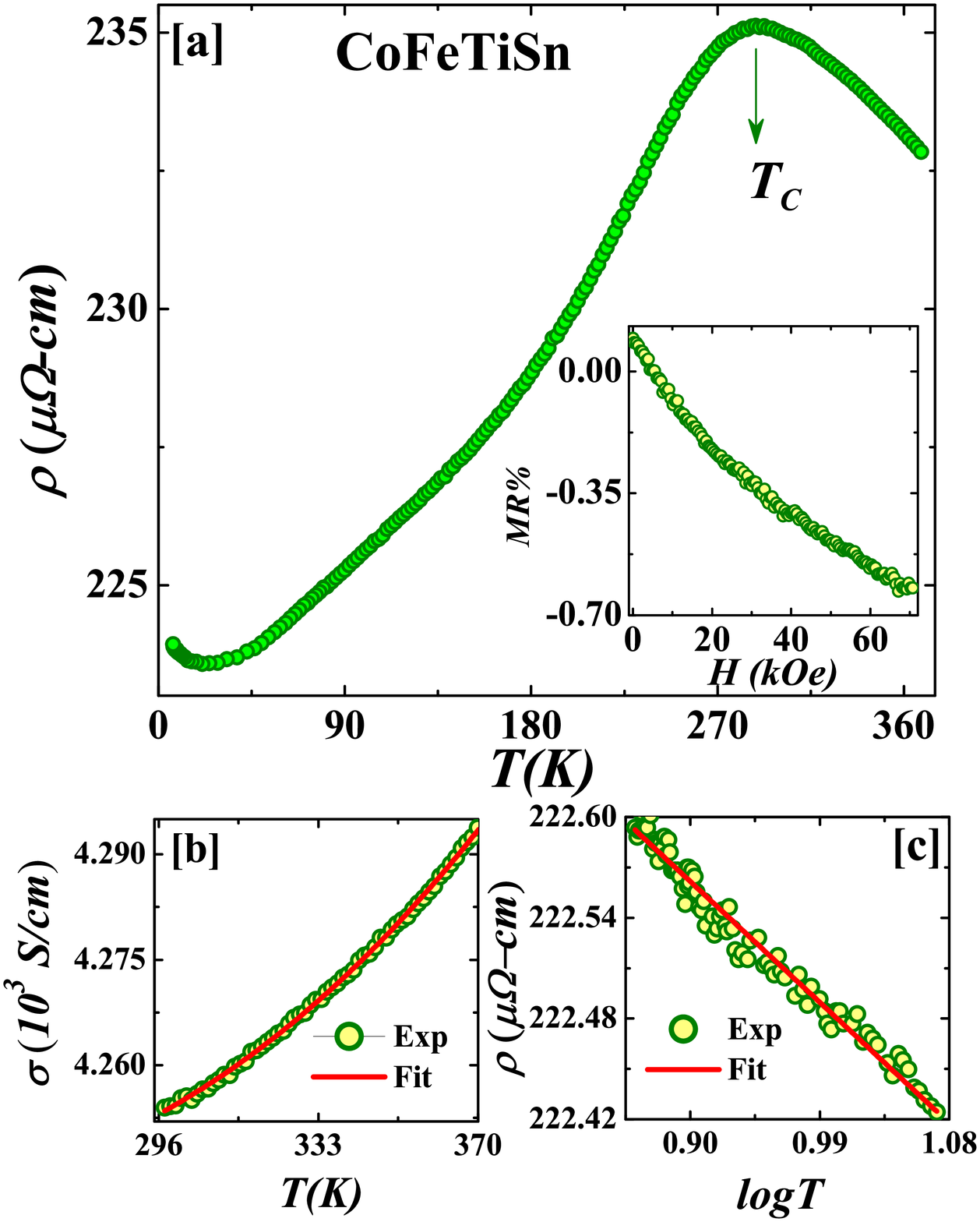}
\caption {The main panel of (a) shows the  $T$ variation of $\rho$ at zero field for CoFeTiSn. The inset shows the magnetoresistance measured at 300 K. (b) shows the conductivity as a function of temperature along with a fitting to the data with the function, $\sigma(T) = \sigma_0 + \sigma_g\exp(-E_g/k_BT)$. (c) shows $\rho$ as a function of $\log{T}$ below 25 K, along with a linear fit to the data.}
\label{rho1}
\end{figure}

\subsubsection{Electrical Resistivity}

The most fascinating observation in the present work is obtained from the transport measurement on the samples. The zero field $T$ variation of $\rho$ of CoFeTiSn is represented in the main panel of fig.~\ref{rho1}(a). It can be seen that the sample shows semiconducting behaviour ($d\rho/dT <$ 0), but with rather low value of resistivity ($\rho \sim$ 220-230 $\mu \Omega$-cm). On cooling, the sample undergoes a semiconductor to metal transition exactly at the FM $T_C$. This indicates that the magnetism is intimately connected with the electronic transport properties of the sample. The electrical conductivity ($\sigma = 1/\rho$) of the high temperature ($> T_C$) semiconducting part can be fitted with a relation $\sigma(T) = \sigma_0 + \sigma_g\exp(-E_g/k_BT)$, where $\sigma_0$ and $\sigma_g$ are $T$-independent coefficients and $E_g$ is the band gap. The fitted data are shown in fig.~\ref{rho1} (b), and the calculated value of $E_g$ is found to be 0.16 eV. Such activated nature of the conductivity indicates the presence of a gap in the electronic states near Fermi level. The value of $E_g$ matches well with other semiconducting Heusler systems~\cite{feval}.

\par
We have also measured $\rho$ under an applied magnetic field. We hardly see any change in $T_C$ under the application of $H$ = 50 kOe. The magnetoresistance (MR = $[\rho(H) - \rho(0)]/\rho(0)$) is found to be negligible around $T_C$ with a negative value [see inset of fig.~\ref{rho1}(a)]. 

\par
Below $T_C$, the sample shows predominantly metallic character, except for the upturn below 25 K. The $\rho(T)$ data below 15 K obeys a logarithmic $T$ variation as evident from the $\rho$ vs.  $\log{T}$ plot in fig.~\ref{rho1} (c). The complete data below 22 K can be well fitted with the relation, $\rho(T) = \rho_0 - \rho_1\log T + \rho_2T^5$, where $\rho_2T^5$ stands for the electron-phonon scattering, $\rho_0$ is the residual resistivity due to impurity and $\rho_1\log T$ is a Kondo-like term (fitting not shown here).     

\par
\begin{figure}[t]
\centering
\includegraphics[width = 7.8 cm]{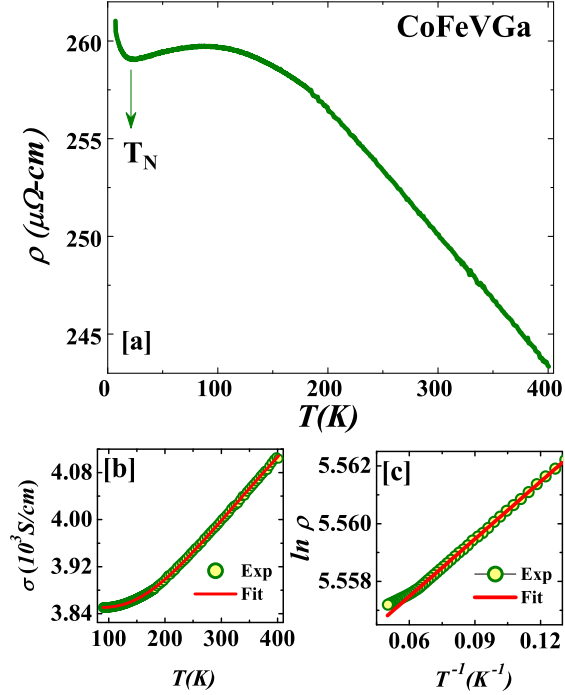}
\caption {(a) shows the $T$ variation of $\rho$ at zero field for CoFeVGa. (b) shows the high temperature conductivity part along with fitting line with the function, $\sigma(T) = A_1\exp{(-E_1/k_BT}) + A_2\exp{(-E_2/k_BT})$ above 90 K. (c) shows the $\log(\rho)$ versus $1/T$ plot below 20 K, along with a linear fitting.}
\label{rho}
\end{figure}

The $\rho(T)$  behaviour of CoFeVGa is found to be significantly different from that of normal metals or semiconductors. The zero field $T$ variation of $\rho$ represented in the fig.~\ref{rho}(a) suggests CoFeVGa is  predominantly a semiconductor ($d\rho$/$dT <$ 0) at least down to 90 K. The sample shows a broad hump like feature in its $\rho(T)$ data around 90 K, and resistivity drops with further lowering of $T$. Such  semiconducting behaviour over a wide temperature window is relatively uncommon in an intermetallic compound. Among Hesuler compounds, Mn$_2$CoAl shows semiconducting behaviour down to 2 K~\cite{felser}, while several quaternary Heusler compounds, such as CoFeCrGa~\cite{bainsla2}, CoFeMnSi~\cite{cofemnsi}, and  CrVTiAl~\cite{venka}, show semiconducting behaviour over a large $T$-range. The quaternary alloys showing such semiconducting behaviour have often been identified as the prospective candidates for SGS state. Whether CoFeVGa belongs to that group requires further investigations. We observe an upturn in $\rho$ below about 20 K, which matches well with the observed AFM-like magnetic transition. The data below 20 K, unlike the  CoFeTiSn sample, shows an activated behaviour, $\rho \sim \exp{(\Delta/T)}$ [see fig.~\ref{rho}(c)]. The value of this activation energy $\Delta$ is found to be 0.067 K.   

\par
The $T$ variation of conductivity of CoFeVGa cannot be fitted with an activated type behaviour with single carrier. On the other hand, if we assume there are two carriers (say electron and holes), we can write $\sigma = e(n_1\mu_1 + n_2\mu_2)$, where $n_i$ and $\mu_i$ are the density and mobility of $i$th carrier. Considering a simple activated type of conduction ($n_i \sim \exp{(E_i/k_BT)}$, the conductivity can be written as $$\sigma = A_1\exp{(-E_1/k_BT)} + A_2\exp{(-E_2/k_BT)}.$$  We fitted the $\sigma(T)$ curve between 90 to 400 K using this two carrier model as shown in fig.~\ref{rho} (b). The best fit is obtained for $E_1$ = 61.15 meV and $E_2$ = 0.035 meV. Interestingly the conductivity value of CoFeVGa at room temperature is 3998 S/cm, which is  much higher than that reported for Mn$_2$CoAl ($\sim$ 2440 S/cm).

\subsubsection{Thermoelectric measurement}

\begin{figure}[t]
\centering
\includegraphics[width = 10 cm]{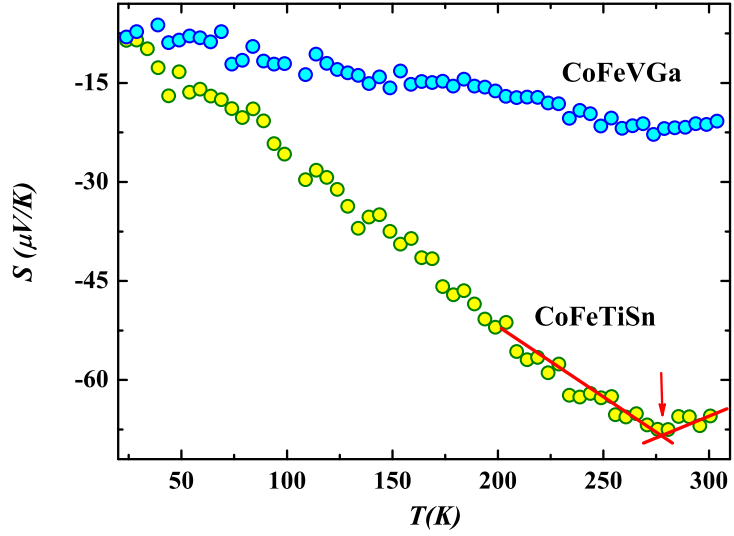}
\caption {$T$ variation of Seebeck coefficient($S$). The solid lines are guide to the eye.}
\label{thermo}
\end{figure}

Thermoelectric studies is an important tool to ascertain the electronic state of a material. Here we measured $T$ dependence of Seebeck coefficient ($S$) for both CoFeTiSn and CoFeVGa  as shown in  fig.~\ref{thermo}. The negative value of $S$ indicates that the electrons are the dominant carrier in these compounds. The maximum values of $S$ for CoFeTiSn and CoFeVGa are found to be around 65 $\mu$V/K and 21 $\mu$V/K, respectively at room temperature. The absolute value of $S$ at 300 K for the above discussed samples are comparable to that of potential thermoelectric Heusler alloy Fe$_2$VAl ($\sim$ 35 $\mu$V/K)~\cite{lue}. Close observation of $S(T)$ data for the sample CoFeTiSn reveals a change in slope at around 280 K. The corresponding temperature, indicated by an arrow in fig.~\ref{thermo}, corroborates with the semiconductor to metal transition as observed in the $\rho(T)$ curve [fig.~\ref{rho1} (a)] and FM $T_C$ [fig.~\ref{mag}(a)] for the CoFeTiSn sample. Such slope change or upturn in $S(T)$curve in high temperature regime is related to the activation of the minority carrier in the system~\cite{A1}. Furthermore, attempt has been made to estimate the Fermi Energy of both the systems from the $S(T)$ data using the relation~\cite{A2}:

\begin{equation}
\centering
S = \pm\frac{k_B}{e}\left[\eta_F - \frac{(r + 5/2)F_{r + 3/2}(d\eta_F)}{(r + 3/2)F_{r + 1/2}(\eta_F)}\right]
\label{ef}
\end{equation}

Here, $\eta_F = \mathcal{E}_F/k_BT$ is the reduced Fermi energy, and $F_n = {\displaystyle \int_0^{\infty}\frac{\eta^n}{1 + \exp{(n-n_F)}}\,d\eta}$. Here $\mathcal{E}_F$ represents energy of the Fermi level measured from the bottom of the conduction band.
Now, $r$ = -0.5 corresponds to acoustic phonon scattering, $r$ = 0.5 is scattering due to optical phonons and $r$ = 1.5 denotes scattering by ionized impurities~\cite{A2}. At room temperature, reasonable values of $\mathcal{E}_F$ viz., 220 meV for CoFeTiSn and 700 meV for CoFeVGa are obtained, provided optical phonon scattering ($r$ = 0.5) dominates in the systems under investigation. DFT calculations on CoFeTiSn indicates that the value of $\mathcal{E}_F$ is between 0.3-0.5 eV for non-spin-polarized and spin-polarized calculations. This value is slightly higher than that estimated from $S$. For  CoFeVGa, DFT value (0.65 eV) is rather close to that estimated from eqn.~\ref{ef}.

\section{Discussion}
The present work addresses the magnetic and the electronic properties of two quaternary Heusler alloys, namely CoFeTiSn and CoFeVGa. CoFeTiSn is found to be primarily metallic below the ferromagnetic $T_C$ of 287 K, while CoFeVGa shows a predominantly semiconducting character. The most fascinating observation on CoFeTiSn is the transition from a semiconducting state to a metallic state at $T_C$. Electronic  structure calculations~\cite{cofetisn} reveals that the compound is metallic as far as the majority carriers are concerned, the valence band of minority carrier just touches the Fermi level, with very small minority spins available for conduction. As a result, the compound can be described as a quasi-half metal with almost 100\% of spin polarization. Our experimental work on CoFeTiSn correspond very well with the previously reported first principle calculation, such as (i) experimental lattice parameter (6.069 \AA) is very close to the theoretical value (6.09 \AA), (ii) the ground state is found to be ferromagnetic as predicted in theory, (iii) the `$Z_t$-24' Slater Pauling rule is not obeyed, although the value of saturation moment is found to be integer.

\par
However, in contrary to the theoretical result, the material shows a semiconducting behaviour (negative $T$ coefficient of $\rho$) in the paramagnetic state. The theoretical density of states (DOS) in the paramagnetic state is associated with a significant population at $E_F$, where $E_F$ lies at the edge of the rising part of the valence band density.  This has been found to be energetically unfavourable, and the system undergoes spin polarization leading to a ferromagnetic state with broader DOS at the $E_F$ for the majority spin. In the literature, there are reports of few intermetallic systems, where semiconductor to metallic transition occurs concomitant with the magnetic ordering. The common experimental evidence of such transition is found in materials such as UNiSn~\cite{unisn}, Co$_2$TiSn~\cite{subham}, doped Fe$_3$Ga and Fe$_3$Si pseudo-binary alloys~\cite{fe3ga,fe3si}, Fe$_2$VAl based Heusler compositions~\cite{nishino}, MnNiGe-based alloys~\cite{prabir}. The semiconductor to metal transition in these systems is generally attributed to one or  combined effect of disorder, large spin fluctuations, small electronic density at $E_F$ and subtle change in electronic states on magnetic ordering. In case of CoFeTiSn, the magnetoresistance near $T_C$ is very small, therefore the role of spin-disorder scattering can be ruled out. The system has $\rho$ value 220 $\mu \Omega$-cm at 4 K, which is large compared to many other intermetallic alloys and compounds. Although, our powder XRD analysis indicates an ordered structure, small antisite disorder may still remain in the sample. However, a sole disorder induced Anderson type localization of carriers seems unacceptable, as the system attains a metallic state below $T_C$.

\par
UNiSn is another widely studied intermetallic compound, where a transition from paramagnetic insulator to AFM semiconductor state is observed. The anomalous transport in UNiSn can be explained on the basis of reconstruction of the bands caused by the polarization dependent exchange interaction of the valence orbitals in presence of  on site Coulomb interaction~\cite{unisn_th}. It is argued that on the onset of AFM transition, the semiconducting gap disappears leading to metallic state. The estimated energy gap is found to be 0.12 eV for UNiSn. A very similar scenario is found in the FM Hesuler compound Fe$_2$VAl, where the system turns metallic from a semiconducting state with the onset of FM order~\cite{nishino}. The semiconducting character of Fe$_2$VAl based alloys has been attributed to the formation of a pseudogap around $E_F$~\cite{djsingh,okamura,nmr}. The spin polarization below $T_C$ destroys the pseudogap, and we see the metallic character.

\par
In case of CoFeTiSn, it is possible that similar pseudogap may originate above $T_C$. The conductivity follows an activated behaviour in the semiconducting state with energy gap close to 0.16 eV, which is very similar to the gap energy of the semiconducting phase of Fe$_2$VAl (0.14 to 0.27 eV) and UNiSi (0.12 eV). The electronic structure calculation on CoFeTiSn indicates it to be metallic in both paramagnetic and FM states. However, the calculation does not take into account the on site Coulomb interaction, $U$.  It is depicted  that $U$ can have significant effect on DOS even in 3$d$ transition metal based Heusler compounds~\cite{sosia}. Therefore, the inclusion of $U$ may open up a gap, which eventually disappears with the onset of magnetic order. CoFeTiSn can be considered as derived from the full Hesler compound F$_2$TiSn, by the replacement of one Fe by Co. The anomalous transport in F$_2$TiSn is found to be related pseudogap in presence of disorder and electron correlation~\cite{fe2tisn}. As already mentioned, CoFeTiSn does not obeys `$Z_t$ - 24' Slater-Pauling rule, although integer value of saturation moment is observed. It may be important to ascertain the electronic state of the material for the implication of  integer moment.

\par
The electronic structure calculation on CoFeVGa claims the sample to be nearly HMF with minority band crossing the $E_F$ a little. The lattice parameter calculated from the first principle is 5.73 \AA, which is very close to our experimental value (5.80 \AA). However, the sample shows a predominantly semiconducting nature between 400-90 K. In addition, CoFeVGa is found to be paramagnetic at least down to 20 K, in sharp contrast with a spin polarized state observed in the calculation. The magnetic ordering takes place at $T_N$ = 20 K, which does not seem to be FM type. The cusp like feature in the ZFC magnetization at $T_N$ may indicate an AFM transition. However, FC magnetization increases below $T_N$ and a clear curvature is seen in the $M-H$ data recorded at 5 K. Notably, $M-H$ curve does not show any tendency of saturation, and therefore, one can rule out the possibility of a pure FM ground state in CoFeVGa. The transition at $T_N$ possibly lead to a canted AFM structure due to the presence of both FM and AFM correlations. It is to be noted that paramagnetic $\theta$ is positive in the system, which indicates the existence of the FM interaction. 

\par
The resistivity of CoFeVGa also shows multiple anomalies. $\rho(T)$ shows a hump like feature around 90 K, The temperature coefficient of resistivity is found to be positive between 90 and 20 K. Below 20 K, the resistivity again shows semiconducting character possibly due to the zone boundary effect associated with canted AFM type ordering. As evident from our result, the $\rho(T)$ data of  CoFeVGa can be fitted with activated two carriers model. We have obtained the band gap for two carriers to be $E_1$ = 61.15 meV and $E_2$ = 0.035 meV. Since the majority carriers are electrons ($S(T) <$ 0), $E_1$ and $E_2$ can be assigned to hole and electrons respectively. 

\par
It is to be noted that CoFeTiSn also shows a low-$T$ upturn below about 25 K, which is not associated with any magnetic anomaly. Unlike CoFeVGa, this upturn cannot be fitted with an activated type behaviour. Rather, the region below 25 K in CoFeTiSn shows a logarithmic contribution in the $T$-dependence of resistivity. Notably, Kondo-lattice behaviour is predicted in the Co and Ni doped Fe$_2$TiSn alloys~\cite{maple}. CoFeTiSn has FM ground state with magnetic ordering taking place close to room temperature. Therefore, it may not be appropriate to associate the low-$T$ upturn in $\rho$ to dense-Kondo effect. We believe that the antisite disorder present in the system can act as magnetic scattering centres, which localize the carriers through dilute Kondo effect.

\par
The Seebeck coefficient is found to be higher in CoFeTiSn than CoFeVGa, despite the fact that the former compound is a metal, while the later compound is a semiconductor. This difference may be due to the complex electronic structure of CoFeTiSn. Being a metal (low value of $\rho$) and having large value of $S$, CoFeTiSn has large value of thermoelectric power factor PF = $S^2/\rho$ of  18.5 $\mu$W.cm$^{-1}$.K$^{-2}$. This value of PF is comparable to good thermoelectric materials like Bi$_2$Te$_3$ and YNiBi~\cite{ynibi}. 

\par
In conclusion, we observe some fascinating transport behaviours in the quaternary Heusler compounds CoFeTiSn and CoFeVGa. Despite the fact that both the compounds has 25 number of valence electrons, there magnetic an d transport properties are found to be significantly different. The semiconductor to metal transition observed around FM $T_C$ has similarity with that of Fe$_2$VAl based Heusler alloys. A psudogap phase, similar to Fe$_2$VAl, can be mooted for the semiconducting phase above $T_C$. CoFeVGa, on the other hand, paramagnetic semiconductor above its magnetic ordering temperature of 20 K, which is in contradiction with the theoretical prediction of a nearly half metallic state. 

\section*{Acknowledgment}
Snehashish Chatterjee would like to acknowledge UGC for his research fellowship. 


\end{document}